# Internal relation between Personality trait Statistical outcomes among Junior College Divers and their performance

CHENG Hua[1]
Sport Science School, Lingnan Normal University
524048, 29# Cunjin Rd, Chikan District, Zhanjiang City, Guangdong Province, China.

## Abstract

Objective: Personality trait can predict divers' behavioral performance underwater. However, we know very little about the innate personality of the junior college diving students. To gain a better insight of personality characteristics of them, we carried out a personality survey base on Eysenck questionnaire. Method: 93 college diving students participated in this survey and totally 74 valid questionnaires recovered. Four dimensions were rating by the self-report scale of 85 questions. Descriptive analysis, *T* test and variance analyses are processing by SPSS20.0. Results: Statistical results indicated that college divers are more extraverted ($t$=10.838, $p$=0.000), more neurotic ($t$=2.747, $p$=0.008) and unlikely psychoticism ($t$=-1.332, $p$=0.187). Differences were found in Gender only in Liar scale score ($F$=7.025, $p$=0.010). Conclusion: These outcomes suggested that parts of the character of the college divers' are suitable for diving activities. And emotional control training is needed in the curriculum setting for them in the process of safe diving.

Keywords： Personality; Extraversion; Neuroticism; Scuba Diving

## Introduction

Personality is believed to be one of the effective index in predicting behavior and task performance of men nowadays [1-5]. Personality characteristics play a key role in training of diving[6]. It certainly constructed important part for mental health for elite athletes including divers [7]. Previous research shows that a significant difference was found among professional divers in nervous activity type, cognitive style, the stability of action, memory span, time reaction, spatial perception, act of attention and dark adaptation[8]. The adapting to environmental stressors for professional divers is complex and requires processing physiologically and psychologically to overcome extreme environmental barriers, thus implement effective and efficient work[9]. Experts have warned us panic behaviors threaten the occupational health and safety of underwater[10]. It was wildly recognized that anxiety could directly lead to panic behavior [11, 12]. And stable personality characteristics help acclimatization to change in an extreme environment. Extraverted and less neurotic athletes in personality are more attracted by the high risk of Sports [13]. Occupation divers in better mental health status are usually higher vocational adaptability than the others. Less anxiety or panic behaviors with good adaptation is of great advantage for the maintenance of security, physical and mental health of divers [4, 5].

---
[1] Email Address: Chengh@lilngnan.edu.cn

Over the past few decades, the self-contained underwater breathing apparatus (scuba) diving has become one of the most popular recreational sport activities in various parts of the world. Scuba diving rapidly becomes the hotspot activity of leisure and competitive sports industry. The need for diving coaches is increasing incredibly nowadays. Since 2013, our sport science school started recruiting high school graduates into scuba diving training course for university education. About 50 students recruited each year for two consecutive years. They were selected at first mostly because their physical, physiological and academic performances. Because of the personality characteristics of them could influence their behavior and safety underwater, mental health and sustainable diving career development as well. Are their personalities traits meet with the characteristics of an excellent diver, or need to be strengthened? This is the problem we want to solve for better sport education during university.

An investigation of personality traits in our sophomore diving students for two consecutive years had been carried out. Focusing on personality characteristics of the diving students will help us find the proper way to cultivate diving talents who ultimately promoting development of the high-end leisure sport industry.

Extraversion is closely aligned with the temperament of positive emotionality or positive affect. And it emerges as a broad dimension in all descriptions of personality structure. Extraversion-introversion contrasts people who are described as sociable, energetic, and assertive with ones who are reserved, withdrawn, and submissive. People who are more extraverted experience greater happiness, subjective and existential well-being than those inclined to introversion. Consistent with an underlying approach temperament, those extraverts are more likely to use coping strategies that involve engaging with a challenge, such as problem-solving, than strategies of disengagement or avoidance. They are more likely to be popular, to get higher social status, to be easily satisfied with their jobs and be accepted by their peers. However, extraversion has only rarely been found to be directly related to some of the other widely studied consequential outcomes such as longevity, marital stability, and occupational success.

The trait of neuroticism is the chronic tendency for some individuals to experience more negative thoughts and feelings than others, to be emotionally unstable, and insecure. In contrast to those who are emotionally stable, more neurotic individuals are prone to be worried, anxious, moody, irritable, and depressed. Neuroticism predicts a wide range of negative outcomes, including psychopathology. People who are more neurotic have lower self-esteem and subjective wellbeing. Higher levels of neuroticism are associated with undesirable interpersonal consequences such as less satisfying relationships and divorce and more aggressive behavior. Neuroticism predicts negative health outcomes such as reporting more somatic symptoms. Recent research confirms the greater emotional reactivity of more neurotic individuals across a variety of everyday settings. Higher neuroticism was associated with stronger appraisal-emotion associations, confirming that, for example, a more neurotic person reacts to unfair events with more anger than a less neurotic person[14].

## Methods

Questionnaire surveys were conducted in underclassman diving students in grade 2013 and 2014 separately in their sophomore year, using *the Eysenck* personality questionnaire(EPQ) of the Chinese version, the self-report scale of 85 questions. Based on the theory of *Eysenck's* trait theory, four dimensions including E (Extraversion), N (Neuroticism), P (Psychoticism) and L (Lie) subscales assess the participants in four different characteristic of personality. Participants choose "yes" or "no" according to their own situation, while experimenter scored them according to the principles. Original points of each subscale score were calculated. Then standard score can be checked in the standard table. The research and analysis of the test results are mainly based on the standard score.

Participants signed informed consent forms before filling out the questionnaires. The experimental process is monitored and all questionnaires entries were unified explained by the experimenter. Participants take their choices after they saw and understood the questionnaire items. They scored themselves after completed the Questionnaires and learned the calculation rules. All initial scores are converted into standard scores according to the statistics table by the experimenter. Participants in Experiments of *Eysenck* questionnaire survey for two grades of diving professional diving college students to collect the data of the sophomores' personality traits. Each time the questionnaires were gathered on the spot. The personality questionnaire test is approved by the university ethics committee.

The average value's standard error of the standard norm is 50±10. There are 68.46％ of normal people between 40 ~ 60 points and 95.45％ of them between 30~70 in our country. If a test standard score greater than 60 or less than 40, it can be considered that the subject has the characteristics of high or low in the scale, if the standard score greater than 70 or less than 30, so these features more obvious.

Collected data were processed by ***IBM SPSS20.0*** for descriptive analysis,

single factor analysis and *t*-test. **Hem I 1.0** software is for the heat map drawing.

# Results

93 participants took part in this survey. The valid response rate is 79.6%, namely entirely 74 valid questionnaires recovered. Those who click all "yes" or "no" in their answers, or have too many missing selection in self-evaluating were excluded. The questionnaire scoring display as in the form of heat map, as shown in *Fig 1*. There are 45 retrieved questionnaires are obtained from Grade 2013 and 29 from Grade 2014, and 63 male and 11 female students had completed these personality questionnaires.

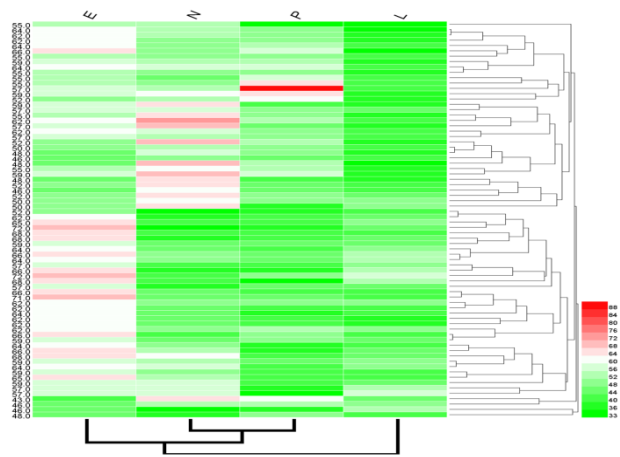

Fig 1  Heat map of EPQ scoring of all diving college students

Average score of subscale-E, N, P and L of the overall sample were made compared with the norm standards (test value= 50). Score in dimension E (*t*=10.838, *p*=0.000) and N(*t*=2.747, *p*=0.008) are higher than the norm. While score in dimension P (*t*=-1.332, *p*=0.187) and L(*t*=-8.036, *p*=0.000) are lower than norm, as showed in *Table 1*.

Table 1  Description Analysis and T test statistics of overall sample

| | E | N | P | L |
|---|---|---|---|---|

| Descriptive analysis | N=74 | $\bar{\chi}\pm s$ | 58.73±6.929 | 53.12±9.777 | 48.64±8.812 | 44.35±6.046 |
|---|---|---|---|---|---|---|
| T-test | Test value = 50 | $t$ | 10.838** | 2.747** | -1.332 | -8.036** |
| | | $p$ | 0.000 | 0.008 | 0.187 | 0.000 |

*\*\*: Significant correlation at p<0.01 (double side)*

30 out of 74 participants (40.5% of all participant) scored over 60 in dimension E, which thought to be are extroverted in characteristic. 17of them (23%) are neurotic basing on their score in Neuroticism. About 82.4% of them scored between 41~60 in Psychoticism. Last but not least in Liar scale. All of them scored below 60, which indicated the test results are trust worthy (*Table 2*).

Table 2　Frequency and percentage in every EPQ scoring intervals of all subjects

| | | Scoring Intervals | E | | N | | P | | L | |
|---|---|---|---|---|---|---|---|---|---|---|
| | | | Frequency | % | Frequency | % | Frequency | % | Frequency | % |
| Overall | | 30~40 | 0 | 0.0 | 8 | 10.8 | 8 | 10.8 | 20 | 27.0 |
| | | 41~60 | 41 | 55.4 | 49 | 66.2 | 61 | 82.4 | 54 | 73.0 |
| | | 61~70 | 30 | 40.5 | 15 | 20.3 | 4 | 5.4 | 0 | 0.0 |
| | | >70 | 3 | 4.1 | 2 | 2.7 | 1 | 1.4 | 0 | 0.0 |
| | | Total | 74 | 100.0 | 74 | 100.0 | 74 | 100.0 | 74 | 100.0 |
| Grade | 2013 | 30~40 | 0 | 0.0 | 5 | 11.1 | 6 | 13.3 | 9 | 20.0 |
| | | 41~60 | 26 | 57.8 | 29 | 64.4 | 36 | 80.0 | 36 | 80.0 |
| | | 61~70 | 17 | 37.8 | 11 | 24.4 | 2 | 4.4 | 0 | 0.0 |
| | | >70 | 2 | 4.4 | 0 | 0.0 | 1 | 2.2 | 0 | 0.0 |
| | | Total | 45 | 100.0 | 45 | 100.0 | 45 | 100.0 | 45 | 100.0 |
| | 2014 | 30~40 | 0 | 0.0 | 3 | 10.3 | 6 | 20.7 | 11 | 37.9 |
| | | 41~60 | 15 | 51.7 | 20 | 69.0 | 21 | 72.4 | 18 | 62.1 |
| | | 61~70 | 13 | 44.8 | 4 | 13.8 | 2 | 6.9 | 0 | 0.0 |
| | | >70 | 1 | 3.4 | 2 | 6.9 | 0 | 0.0 | 0 | 0.0 |
| | | Total | 29 | 100.0 | 29 | 100.0 | 29 | 100.0 | 29 | 100.0 |
| Gender | M | 30~40 | 0 | 0.0 | 6 | 9.5 | 9 | 14.3 | 19 | 30.2 |
| | | 41~60 | 35 | 55.6 | 42 | 66.7 | 50 | 79.4 | 44 | 69.8 |
| | | 61~70 | 27 | 42.9 | 13 | 20.6 | 4 | 6.3 | 0 | 0.0 |
| | | >70 | 1 | 1.6 | 2 | 3.2 | 0 | 0.0 | 0 | 0.0 |
| | | Total | 63 | 100.0 | 63 | 100.0 | 63 | 100.0 | 63 | 100.0 |
| | F | 30~40 | 0 | 0.0 | 2 | 18.2 | 3 | 27.3 | 1 | 9.1 |
| | | 41~60 | 6 | 54.5 | 7 | 63.6 | 7 | 63.6 | 10 | 90.9 |
| | | 61~70 | 3 | 27.3 | 2 | 18.2 | 0 | 0.0 | 0 | 0.0 |
| | | >70 | 2 | 18.2 | 0 | 0.0 | 1 | 9.1 | 0 | 0.0 |
| | | Total | 11 | 100.0 | 11 | 100.0 | 11 | 100.0 | 11 | 100.0 |

## Grade differences

Statistically comparisons were made between Grade 2013 and the Grade 2014. The description of EPQ scoring and the frequency and percentage in scoring intervals of the Two grades has set out in *Table 3*, *figs 2 & 3*. Test score in dimension E, N, P and L respectively are 58.64±7.22 *vs.*58.86±6.58, 53.22±9.70 *vs.*52.97±10.06, 48.8±9.55 *vs.* 48.38±7.68, 44.78±5.83 *vs.* 43.69±6.42. Single factor analysis was conducted by grade as an independent variable with each subscale scoring as the dependent variables. The differences between grades are not significant statistically, as showed in *Table 3*. It suggested that participant from two grades are very close.

Table 3  Grade related statistics analysis of EPQ scoring

|  |  |  | E | | N | | P | | L | |
|---|---|---|---|---|---|---|---|---|---|---|
|  |  |  | 2013 | 2014 | 2013 | 2014 | 2013 | 2014 | 2013 | 2014 |
| Descriptive analysis | $n$ | | 45 | 29 | 45 | 29 | 45 | 29 | 45 | 29 |
| | $\bar{\chi}$ | | 58.64 | 58.86 | 53.22 | 52.97 | 48.8 | 48.38 | 44.78 | 43.69 |
| | $s$ | | 7.218 | 6.578 | 9.704 | 10.059 | 9.553 | 7.678 | 5.827 | 6.42 |
| Single factor analysis | $F$ | | 0.017 | | 0.012 | | 0.04 | | 0.568 | |
| | $p$ | | 0.896 | | 0.913 | | 0.843 | | 0.454 | |
| A single sample $t$-test | Standard value =50 | $t$ | 8.034 | 7.255 | 2.227 | 1.588 | -0.843 | -1.137 | -6.012 | -5.293 |
| | | $p$ | 0.000** | 0.000** | 0.031* | 0.124 | 0.404 | 0.265 | 0.000** | 0.000** |

*$p<0.05$；** $p<0.01$.*

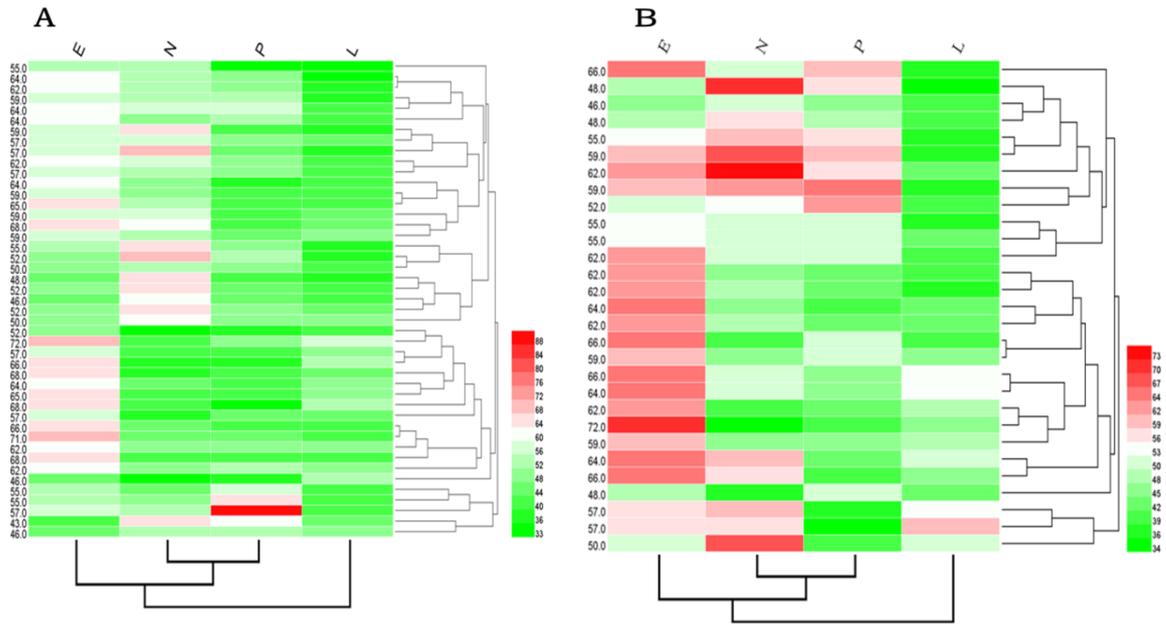

Fig 2　Heat maps of EPQ scoring in different Grades

A: heat map of EPQ scoring in **Grade 2013**；B: heat map of EPQ scoring in **Grade 2014**.

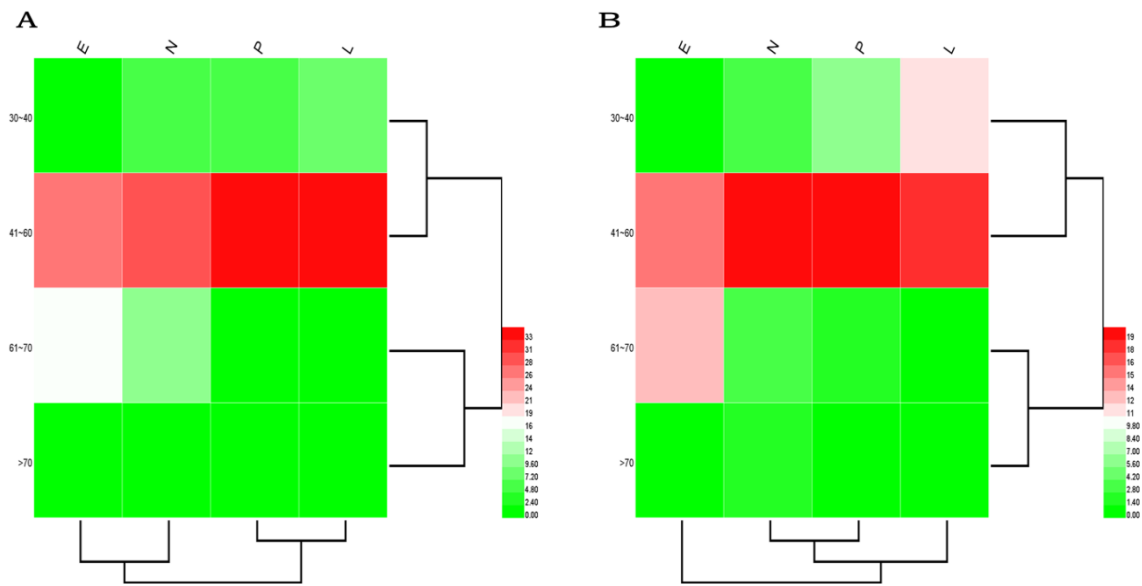

Fig 3　Heat maps of EPQ scoring intervals frequency in different Grades

A: heat map of EPQ scoring intervals frequency in **Grade 2013**；B: heat map of EPQ scoring intervals frequency in **Grade 2014**.

As the whole diving college sample, the scores are higher than normal in EPQ-E and EPQ-N while lower in scoring of EPQ-P and EPQ-L both in grade 2013 and grade 2014，when comparing was made between the diving students and the norm (test value =50). The disparity between diving students and normal is significant in EPQ-E ($t$=8.034, 7.255, $p$=0.000) and EPQ-L scoring($t$=-6.012, -5.293, $p$=0.000) as showed in *Table 3*.

**Gender disparity**

Statistically analyses were further compared between Male and Female diving college students. EPQ scoring and the frequency and percentage in scoring intervals of different gender have shown in *Table 2*, *figs 4&5*. Disparity of average and standard deviation in scoring dimension E (58.59±6.703 *vs.* 59.55±8.43), N (53.57±9.7*vs.* 50.55±10.289), P (48.6±7.228*vs.* 48.82±15.587) in Male *vs.* Female were not significant. Single Factor analyses indicated that female diving college students are more likely to lie to fit in, as in dimension L scoring (43.6±5.712*vs.* 48.64±6.392) in *Table 4*. Gender has a statistically significant impact only on Liar scale score($F$=7.025, $p$=0.010).

Table 4 Gender related statistics analysis of EPQ scoring

|  |  |  | E | | N | | P | | L | |
|---|---|---|---|---|---|---|---|---|---|---|
|  |  |  | Male | Female | Male | Female | Male | Female | Male | Female |
| Descriptive analysis | $n$ | | 63 | 11 | 63 | 11 | 63 | 11 | 63 | 11 |
|  | $\bar{x}$ | | 58.59 | 59.55 | 53.57 | 50.55 | 48.6 | 48.82 | 43.6 | 48.64 |
|  | $s$ | | 6.703 | 8.43 | 9.7 | 10.289 | 7.228 | 15.587 | 5.712 | 6.392 |
| Single factor analysis | $F$ | | 0.177 | | 0.896 | | 0.005 | | 7.025 | |
|  | $p$ | | 0.675 | | 0.347 | | 0.941 | | 0.010* | |
| A single sample $t$-test | Standard value =50 | $t$ | 10.169 | 3.755 | 2.922 | 0.176 | -1.534 | -0.251 | -8.888 | -0.708 |
|  |  | $p$ | 0.000** | 0.004** | 0.005** | 0.864 | 0.13 | 0.807 | 0.000** | 0.13 |

*$p$<0.05；**$p$<0.01.

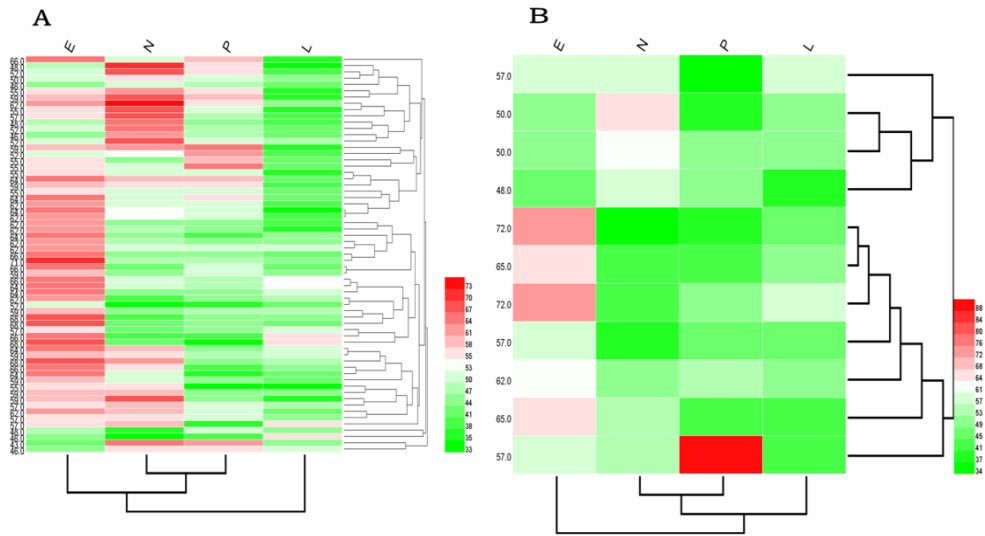

Fig 4　Heat maps of EPQ scoring in different Gender

A: heat map of EPQ scoring in **male;** B: heat map of EPQ scoring in **female**.

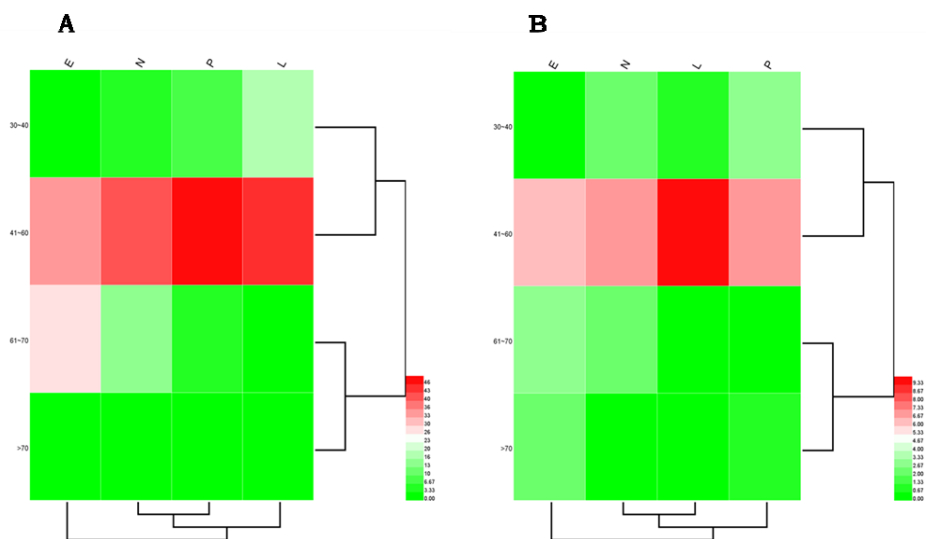

Fig 5　Heat maps of EPQ scoring intervals frequency in different

A: heat map of EPQ scoring intervals frequency in **male**；B: heat map of EPQ scoring intervals frequency in **females**.

Both female and male diving students are more extraverted as comparing with the normal. But male diving students are inclined to be more emotionally unstable. At the level of $p<0.05$, personality distinctions in EPQ-E、EPQ-N and EPQ-L are statistically significant.

**Grade and gender correlation**

Pared-samples *T* test was established between participant in different grade and gender at every scoring interval. High correlation is available for consultation in EPQ-E, EPQ-N, EPQ-P and EPQ-L subscales between two grades. And the analogous situation was observed in gender, too. At the level of $p < 0.05$, correlations among Grades and Genders were statistically significant (See *Table 5*).

Table 5    Relationships between Grades and Genders in EPQ scoring intervals

| Impact factor | Grade | | | | | | | | Gender | | | | | | | |
|---|---|---|---|---|---|---|---|---|---|---|---|---|---|---|---|---|
| Subscale | E | | N | | P | | L | | E | | N | | P | | L | |
| | 2013 | 2014 | 2013 | 2014 | 2013 | 2014 | 2013 | 2014 | M | F | M | F | M | F | M | F |
| *R* | 0.984 | | 0.970 | | 0.996 | | 0.937 | | 0.902 | | 0.925 | | 0.952 | | 0.943 | |
| *p* | 0.002** | | 0.006** | | 0.000** | | 0.019* | | 0.036* | | 0.024* | | 0.013* | | 0.016* | |

*M: male; F: female.*

*\* p<0.05；\*\* p<0.01*

## Discussions

### Grade differentiation

The distribution of students' numbers in each scoring interval of every subscale is quite similar in two grades. 57.8%~80% in Grade 2013 and 54.5%~72.4% in Grade 2014 are scoring 41~60. Students with propensity and extreme personality traits are few or minimal. Students in 2 grades were more extroverted, less concealing than normal crowd. Students from grade 2013 are more likely emotionally unstable. Overall sample participant is tested at similar age and sharing the same educational background. Few differences are found in these 2 grades. Age trajectories for personality traits are similar in this college divers group, which was in line with anticipation.

**Gender differentiation**

55.6%~79.4% of male and 54.5%~90.9% of female diving students are scoring 41~60. Both male and female diving students are extroverted and emotionally unstable with less concealing. Male students are much more sincerity than girls.

Investigation results indicated that extroverted, neurotic and less concealing are the personality characteristics of diving students. Extroversion means being talkative, bold, spontaneous, sociable, dominant and energetic [15]. Extroversion tendency is suitable for high risk and challenging activities, like diving *per se*. Neuroticism is related to stress and depression [16]. It refers to the relatively stabilized responding tendencies of negativity to threat, frustration or loss [17]. It is often used by mental and general health service to predict mental and physical disorders and their comorbidity. Many psychologists also thought neuroticism is a predictor of the quality and longevity of our lives. High neuroticism in personality traits is obviously not conducive to the underwater environment stress adaptation.

All investigated participant show no psychotic symptoms in psychoticism dimension. And the results of the investigation are genuine and believable according to the scoring in Liar subscale.

According to the data, psychological adjustment ability training for strengthening emotional stability and cooperation consciousness and skills coaching for mutual assistance when necessary should be compulsory course for the diving training, especially in curriculum.

## Perspectives

Extraversion and neuroticism are related to different emotions according to the consensus of academic. They have a positive effect on affective states mediated by cognitive appraised and influenced one's affective state. Diving triggers physiological stress response in sinus bradycardia, vasoconstriction, and decrease in effective circulating blood volume and an increase in peripheral circulatory perfusion[18]. Extroverts tend to engage in high-risk activities. The statistical result of the study suggests that extraversion trait might be benefiting the diving underclassmen in diving performance. Neuroticism is not conducive to strategic adjustment and emergency response in sport. Unstable coping behaviors endanger divers' safety underwater. In this survey, we also confirm that gender plays an important role in personality of diving athletes. The outcome is similar with some previous researches[19, 20]. Because of concerning too much about their body shape, high power-to-weight ratio, and sport utilizing weight categories, female diving athletes are more likely to be in poor psychological outcomes[21] . The facts we mentioned telling that personality trait can be a problem in the athletic diving community and curriculum design for coaching and training regarding this challenge are important.

## Acknowledgements

Special thanks are for Zhanjiang Diving School that provides experimental area and the main experimental equipment. And also like to thank the diving instructors from Zhanjiang Diving School who had offered me a lot of help and in the experiment.
This study is funded by①Science Research Project of Lingnan Normal University，《Study on the

Relationship between the Asymptomatic VGE and VA/Q in single no-decompression dive》（ZL1508）;② Research Program of Science and Technology of Zhanjiang《Pretreatment of HBO vs. NBO for Intervention of VGE in Air Diving: a Randomized Double-blind Controlled Study》（2015B01115）.

## Disclosure statement

No potential conflict of interest was reported by the author.